\documentclass[twocolumn,preprintnumbers,amsmath,amssymb]{revtex4}

\usepackage{amsmath}    % need for subequations
\usepackage{graphicx}   % for figures
\usepackage{color}

\begin{document}

\title{A remark on the role of indeterminism and non-locality in the violation of Bell's inequality}
\author{Massimiliano Sassoli de Bianchi}
\affiliation{Laboratorio di Autoricerca di Base, 6914 Carona, Switzerland}\date{\today}
\email{autoricerca@gmail.com}   %optional

\begin{abstract}

Some years ago Aerts \emph{et al.}~\cite{AertsBroekaert} presented a macroscopic model in which the amount of non-locality and indeterminism could be continuously varied, and used it to show that by increasing non-locality one increases, as expected, the degree of violation of Bell's inequality (BI), whereas, more surprisingly, by increasing indeterminism one decreases the degree of the violation of BI. In this note we propose a different macroscopic model in which the amount of non-locality and indeterminism can also be parameterized, and therefore varied, and we find that, in accordance with the model of Aerts \emph{et al.}, an increase of non-locality produces a stronger violation of BI. However, differently from their model, we also find that, depending on the initial state in which the system is prepared, an increase of indeterminism can either strengthen or weaken the degree of  violation of BI.\\

%\keywords{Quantum-machines \and Entanglement \and Bell's inequalities}

\end{abstract}

\maketitle

\section{Introduction}
\label{Introduction}

In~\cite{Aerts-rod} Aerts constructed an remarkable macroscopic model in which he could operationally define coincidence experiments violating (the CHSH version of) Bell's inequality (BI), with exactly the same numerical value $2\sqrt{2}$ as the one obtained in typical coincidence experiments with entangled microscopic entities in a singlet state~\cite{Aspect1, Aspect2} (which corresponds to the maximal violation obtainable in quantum mechanics~\cite{Cirel}). Subsequently, the model was generalized in~\cite{AertsBroekaert}, with the introduction of two parameters, $\epsilon,\rho\in[0,1]$, quantifying the degree of indeterminism and of non-locality present in the model, respectively. 

More precisely, $\epsilon =0$ corresponds to the classical situation of absence of indeterminism, whereas $\epsilon = 1$ to the situation of maximum indeterminism, typical of pure quantum systems. On the other hand, $\rho=0$ corresponds to the situation of maximum locality, when the two pairs forming the double-system are totally disconnected, whereas $\rho=1$ corresponds to the opposite situation of perfect correlation. 

The authors of~\cite{AertsBroekaert} obtain that BI can only be violated if $\rho\neq 0$, and that the violation takes its maximal numerical value $4$ when $\rho=1$ (maximum correlation) and $\epsilon =0$ (minimum indeterminism). Also, they find that for any $\rho\leq 1/\sqrt{2}$, it is always possible to restore the validity of BI by increasing $\epsilon$, whereas this is not any more possible if $\rho>1/\sqrt{2}$.

To sum up, the study of the model described in~\cite{AertsBroekaert} has showed that the source of the violation of BI is the existence of a non zero correlation between the two pairs forming the double-system ($\rho\neq 0$), whereas the only effect of increasing the level of indeterminism (increasing $\epsilon$) is to decrease the value the inequality can take. 

The purpose of the present paper is to analyze a different macroscopic model in which two parameters $\epsilon$ and $\rho$ will also be introduced, as in~\cite{AertsBroekaert}, to continuously vary the level of indeterminism and non-locality (correlation). This will allow us to confirm that a non zero correlation ($\rho\neq 0$) is the necessary condition for the violation of the inequality. However, we will also show that by increasing the indeterminism (increasing $\epsilon$) we can either increase or decrease the value BI can take, depending on the state in which the system is prepared before the coincidence experiments.

\section{Bell's inequality}
\label{Bell's inequality}

Before describing our macroscopic model, we  briefly recall the expression of Bell's inequality (BI).~\cite{Bell0, Bell1}. On a given physical entity, we assume that four different experiments can be performed: $e^A_a$, $e^A_{a'}$, $e^B_b$ and $e^B_{b'}$. We call $o^A_a$, $o^A_{a'}$, $o^B_b$ and $o^B_{b'}$ the outcomes associated to these four experiments, which can only take the values $+1$ or $-1$. We also assume that experiments $e^A_a$ and $e^A_{a'}$ can be performed together with either of experiments $e^B_b$ and $e^B_{b'}$, so defining additional \emph{coincidence} experiments: $e_{ab}^{AB}$,  $e_{ab'}^{AB}$, $e_{a'b}^{AB}$ and $e_{a'b'}^{AB}$. To every coincidence experiment $e_{cd}^{AB}$, $c\in\{a,a'\}$, $d\in\{b,b'\}$, we can then associate the expectation value $E^{AB}_{cd}$ of the product of outcomes $o^A_c o^B_d$, by:
\begin{eqnarray}
\label{expectation value}
E^{AB}_{cd}&=&\sum {\cal P}_{cd}^{AB}(o^A_c,o^B_d) o^A_co^B_d\nonumber\\
&=& +{\cal P}_{cd}^{AB}(+1,+1) + {\cal P}_{cd}^{AB}(-1,-1)\nonumber\\
&\phantom{=}& - {\cal P}_{cd}^{AB}(+1,-1) - {\cal P}_{cd}^{AB}(-1,+1),
\end{eqnarray}
where ${\cal P}_{cd}^{AB}(o^A_c,o^B_d)$ is the probability that the coincidence experiment $e_{cd}^{AB}$ yields the outcomes $(o^A_c,o^A_d)$. 

Under certain hypothesis (usually referred to as \emph{Bell locality}, which have to do with the existence of hidden variables independently determining the experiments' outcomes), one can prove the following (Bell) inequality~\cite{Bell1, Bell0}:
\begin{equation}
\label{Bell inequalities}
I\equiv |E^{AB}_{ab} - E^{AB}_{ab'}| + |E^{AB}_{a'b'} + E^{AB}_{a'b}|\leq 2.
\end{equation}

Inequality (\ref{Bell inequalities}) is generally violated by quantum systems in entangled states, like for instance those formed by two spin-$1/2$ entities in a \emph{singlet (zero) spin state}, for which one can show that $I= 2\sqrt{2}>2$.~\cite{Aspect1, Aspect2} This means that entangled quantum systems generally violate Bell's locality assumption, a fact which remains true even when the two subsystems are separated by a very large spatial distance. In other terms, no local physical theory in the sense specified by Bell can agree with all statistical implications of quantum mechanics, and spatial separation is not a sufficient condition for \emph{experimental separation}.

\section{A rolling prism}
\label{A rolling prism}

\begin{figure}[!ht]
\centering
\includegraphics[scale =.4]{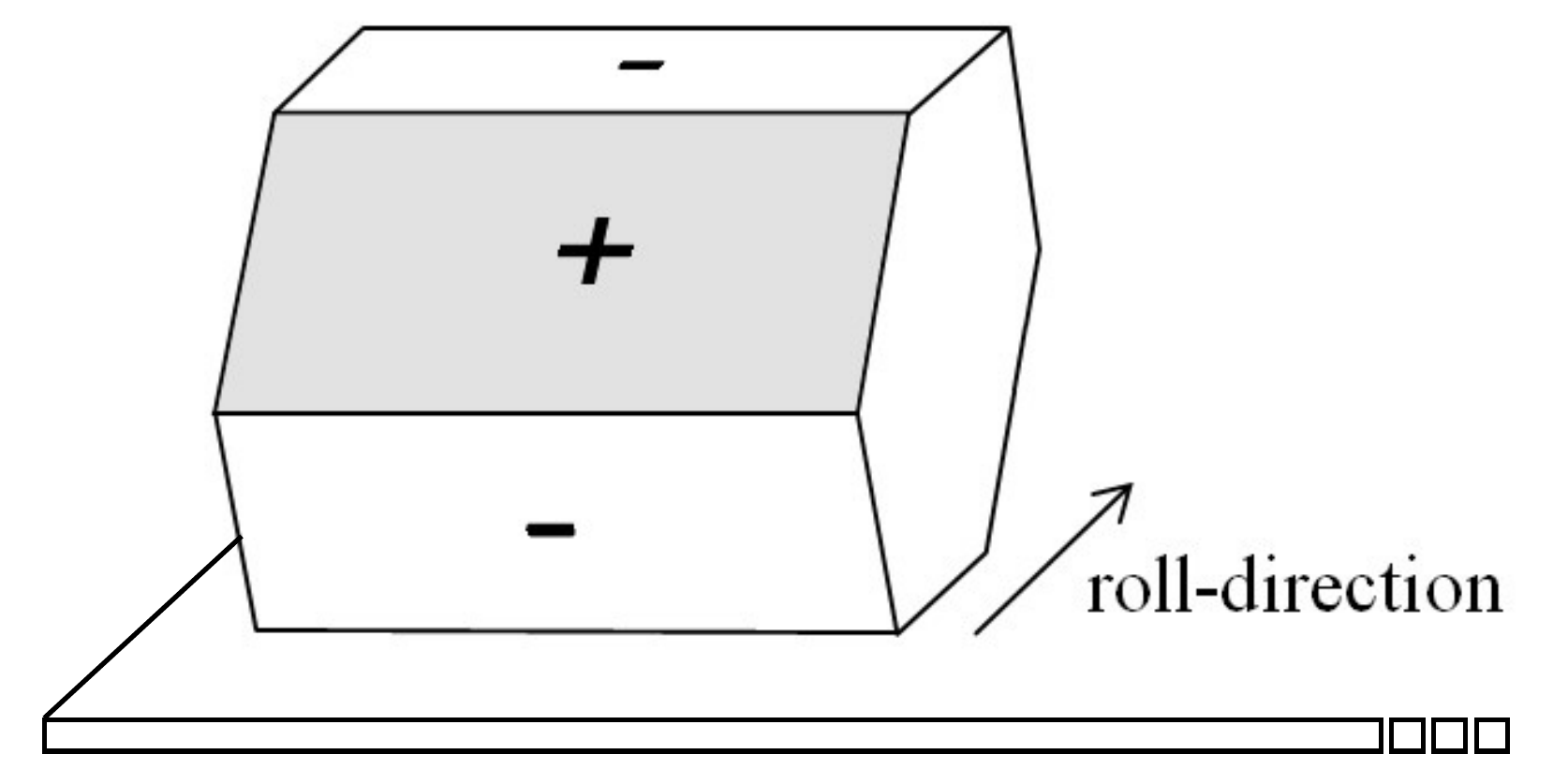}
\caption{An hexagonal prism, placed on a flat surface on one of its six rectangular faces, showing here a ``-'' upper face.
\label{esagono singolo}}
\end{figure}

Let us consider a solid object, of homogeneous density, shaped as a rectangular $n$-prism ($n$ even), i.e., as a polyhedron formed by $n$ identical parallelogram-faces, and two lateral regular polygon-faces with $n$ sides (see Fig.~\ref{esagono singolo} for the case $n=6$, of an hexagonal prism). 

We consider the prism as a special kind of die with $n$ faces, which can be rolled on a flat surface (perpendicular to the gravitational field), along a given ``roll-direction,'' parallel to the two short sides of the rectangular upper face (see Fig.~\ref{esagono singolo}). Each rectangular face of the prism shows either a symbol ``$+$,'' or a symbol ``$-$.'' The ``+'' symbol is only printed on $2$ parallel (opposed) rectangular faces of the prism, whereas the symbol ``-'' is printed on all the remaining $n-2$ rectangular faces. 

We also assume that the ``+'' and ``-'' faces are made of different materials. The two ``+'' faces can slide with an extremely low friction on the flat surface, whereas the $n-2$ ``-'' faces present a very high coefficient of friction with respect to it. Considering that the upper and lower faces of the prism always present the same symbol (and therefore are made of the same material), we have that when the prism presents a ``+'' upper face, it can slide with almost no friction, whereas it cannot do so if the upper face has a ``-'' symbol. 

\begin{figure}[!ht]
\centering
\includegraphics[scale =.5]{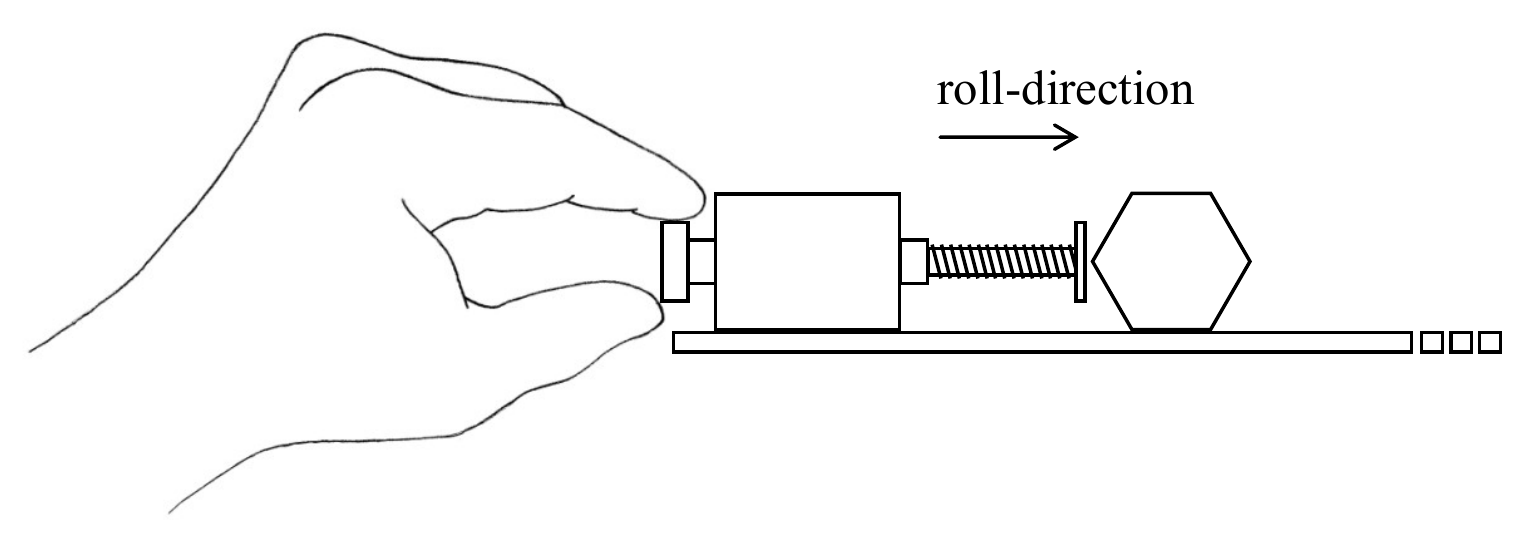}
\caption{In a rolling experiment the experimenter hits strongly the prism (here with six rectangular faces) with the shooter, along the indicated roll-direction, causing it to roll or slide, according to its initial state.
\label{shooter esagono}}
\end{figure}

In the following, when the prism presents a ``+'' (respectively, ``-'') upper face, we shall simply say that it is in state (+) [(respectively, state (-)]. Let us now describe what we shall call a \emph{rolling experiment} with the prism. It consists in placing a specific shooter (similar to a ``flipper ball shooter'') behind the prism, along the roll-direction, pulling firmly its knob and then releasing it, thus communicating to the prism an a priori unpredictable impulsion (see Fig.~\ref{shooter esagono}). One then waits until the prism stops completely, and read the symbol on its upper face. If it is a ``+','' the outcome of the experiment is the value $+1$, otherwise the value $-1$. 

If the prism is in state $(+)$, then, because of the very low friction of the face in contact with the surface, it will not roll, but only glide on it, until all translational kinetic energy will be converted into heat. Therefore, the rolling experiment will not change the prism's upper face, and the outcome will be $+1$, with certainty. 

On the other hand, if the prism is in state $(-)$, then the face in contact with the surface will present an extremely high coefficient of friction with it, so that the prism will not anymore slide but roll (i.e., rotate around its longitudinal axis).  Typically, most of the energy communicated to it by the shooter will be initially transformed into rotational kinetic energy, then, because of the positive work performed by the friction forces, the rotational energy will be gradually transformed into translational kinetic energy and heat, and of course in the end the prism will stop and show a specific upper face. However, the prism will roll only for as long as the non-elastic effects associated with the \emph{rolling frictions} remain lower than the \emph{sliding frictions}, since in this case the prism requires less energy to be moved by rolling than by sliding. But since two of the $n$ faces involved in the rolling movement (those with the ``+'' symbol) present a very low sliding friction, it is highly probable that the prism will conclude its run sliding on one of them, before it will ultimately totally stops. 

In other terms, apart from exceptional circumstances, which we can simply ignore not to complicate unnecessarily our discussion, we can \emph{ideally} assume that, following a rolling experiment, if the initial state is $(-)$ then the final state will be $(+)$, with certainty. So, our prism is so conceived that, independently of its initial state [$(+)$ or $(-)$], the outcome of a rolling experiment will always be $+1$ [i.e., its final state will always be $(+)$].

\section{A double-prism system}
\label{A double-prism system}

\begin{figure}[!ht]
\centering
\includegraphics[scale =.45]{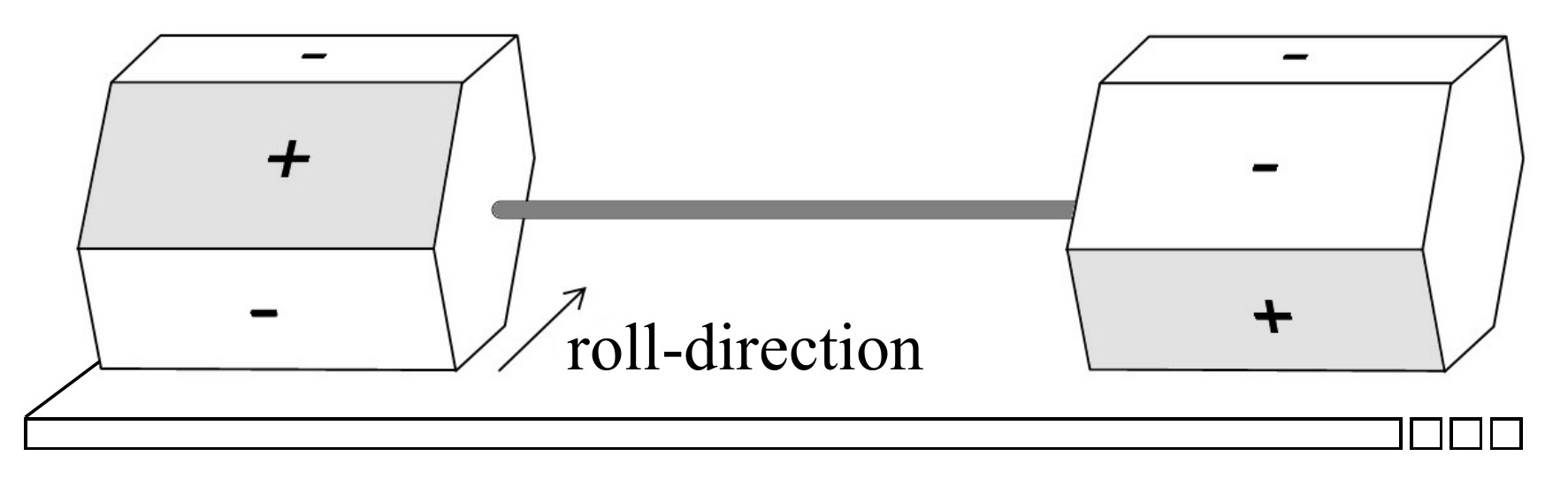}
\caption{A double-system formed by two identical prisms (here with six rectangular faces) connected through space by a rigid rod, glued on their two opposed polygon-faces.
\label{esagoni-connected}}
\end{figure}

We now consider two identical $n$-prisms, and use them to construct an entangled system by connecting them through space, by means of a rigid rod (the length of the rod is arbitrary, but we assume it is made of an extremely light and rigid material) whose two ends are glued at the center of the two opposed polygon-faces of the two prisms, as indicated in Fig.~\ref{esagoni-connected}. 

\begin{figure}[!ht]
\centering
\includegraphics[scale =.4]{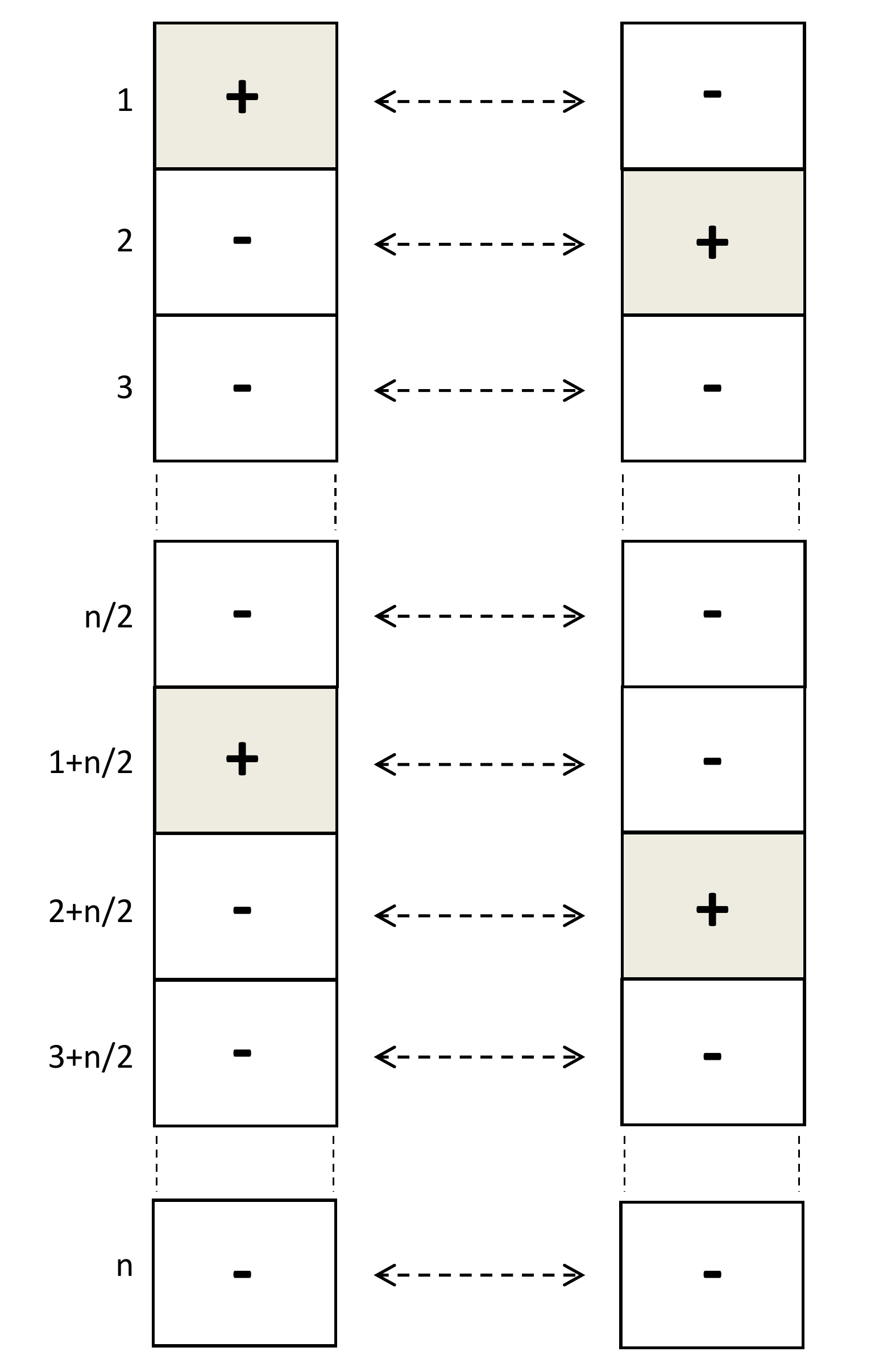}
\caption{The correlations between the $n$ rectangular faces of the two prisms, because of the presence of the connecting rod. 
\label{n-prism-correspondence}}
\end{figure}

Clearly, the connecting rod creates correlations between the $n$ different rectangular faces of the two prisms. Here we assume that the two prisms have been connected in such a way that the correlations in question are those described in Fig.~\ref{n-prism-correspondence}.

We assume that the glue used to connect the rod to the two prisms is strong enough, so that if the two prisms are hit together, simultaneously (in the same direction) they will be able to maintain their connection while rolling, i.e., to remain a whole entity. But we also assume that the glue, although strong, is not as strong as to allow the two prisms to remain connected if only one prism is hit at a time.   

In other terms, if we hit with the shooter only one of the two prism, then, because of the inertia of the other prism, the impact will cause the rod to suddenly detach and fall, thus disconnecting the two solids (one of which will then move in the roll-direction, whereas the other one will remain still). On the other hand, if two experimenters hit with two shooters both prisms at the same time, then the torque experienced by the rod will be much lower, so that the latter will not detach and the two prisms will be able to roll together on the surface, as a one piece entity.

\section{Violating Bell's inequality}
\label{Violating Bell's inequality}

On the double-prism system we have just defined, we perform four different coincidence experiments and show that they produce a violation of BI. For this, we assume that two experimenters (who we shall call experimenter $A$ and experimenter $B$) are placed close to each one of the two prisms. Experimenter $A$ can perform on his prism (say, the left one) two different experiments, $e^A_a$ and $e^A_{a'}$, which are defined as follow. 

Experiment $e^A_a$ is the rolling experiment we have defined in Sec.~\ref{A rolling prism}: it consists in hitting the left prism with a shooter along the roll-direction, then reading the symbol marked on the obtained upper face, producing in this way one of the two outcomes: $o^A_a=+1$, or $o^A_a=-1$. Experiment $e^A_{a'}$ is much simpler: it simply consists in looking at the prism's upper face and check whether it is flat or not. If it is so, then the outcome is $o^A_{a'}=+1$, otherwise it is $o^A_{a'}=-1$. 

Experimenters $B$ can perform on its prism (the right one) the same two experiments as experimenters $A$. In other terms, $e^B_b$ is defined as $e^A_a$, and $e^B_{b'}$ as $e^A_{a'}$. 

Clearly, all of the four above mentioned experiments, when singly performed, can only produce the outcome $+1$. Also, when the coincidence experiments $e_{ab'}^{AB}$, $e_{a'b}^{AB}$ and $e_{a'b'}^{AB}$ are performed by the two experimenters, the only possible outcome for them is $(+1,+1)$, so that according to (\ref{expectation value}), $E^{AB}_{ab'}= E^{AB}_{a'b}=E^{AB}_{a'b'}=1$.

The situation is however more articulate for experiment $e_{ab}^{AB}$, which creates correlations between upper faces. Indeed, if the two experimenters hit simultaneously the two prisms, then, as we explained, the rod will not separate and they will remain connected as they roll (the double-prism system cannot slide, but only roll, as one of its two lower faces is always a high-friction face). 

Therefore, according to Fig.~\ref{n-prism-correspondence}, we have that the probability for the outcome $(-1,-1)$ is zero, the probability for each one of the two outcomes $(-1,1)$ and  $(1,-1)$ is $2/n$, and the probability  for the outcome $(-1,-1)$ is $(n-4)/ n$. Thus, $E^{AB}_{ab} = 0 -(2/n) -(2/n)+[(n-4)/n] = 1-(8/n)$. Inserting all this in (\ref{Bell inequalities}), we obtain: 
\begin{equation}
\label{violation-n}
I = |1-\frac{8}{n}-1|+|1+1| = 2+\frac{8}{n}.  
\end{equation}

Eq. (\ref{violation-n}) clearly violates BI (\ref{Bell inequalities}). It does so in a maximal way ($I=4$) for the case $n=4$ of two tetragonal prisms, and with a value which is very close to the quantum mechanical maximum of $2\sqrt{2}\approx 2.83$ for the case $n=10$ of two decagonal prisms ($I=2.8$). 

Before continuing in our analysis, let us mention that it was Diederik Aerts who was the first, in the early eighties of last century, to conceive an explicit macroscopic model with non-local correlations violating BI in a maximal way,~\cite{Aerts82, Aerts5}, as our double-prism system with $n=4$ can do (see also Ref.~\cite{Massimiliano-elastic} and the references cited therein). 

An important difference between our double-prism model and the historical (connected vessels of water) model of Aerts, is that in the latter it is the fact that the system is broken which is the mechanism responsible for the creation of correlations, whereas in our model it is exactly the contrary: it is only when the double system is \emph{not} broken that the coincidence experiment $e_{ab}^{AB}$ can produce correlations between upper faces (it is important to distinguish ``faces'' from ``upper faces,'' as only the latter are created in a rolling experiment: if we don't create upper faces, by rolling the prisms, BI cannot be violated~\cite{Aerts2, Massimiliano-dice, Massimiliano-elastic}).

\section{Parameterizing indeterminism and non-locality}
\label{Parameterizing indeterminism and non-locality}

We want now to vary the amount of indeterminism and non-locality in our model, to see how this can affect the degree of violation of BI, thus elucidating their role in the violation. First of all, let us observe that if the number $n$ of rectangular faces of the two prisms tends to infinity, the probability ${\cal P}_{ab}^{AB}(-1,-1)= (n-4)/n$ tends to $1$, so that in this limit, according to (\ref{violation-n}), $I\to 2$. 

One could be tempted then to affirm that the level of indeterminism decreases as $n\to\infty$, and that this produces a corresponding decrease of the degree of violation of BI, contrary to what was obtained by Aerts and collaborators, who showed instead that a decrease of indeterminism produces a stronger violation of BI.~\cite{AertsBroekaert}

This however would be a wrong conclusion. Indeed, if by increasing $n$ we can reduce the value taken by BI, this is so because for each different $n$ we have a different physical system, and not because we are decreasing the level of indeterminism in a given physical system. What we have to do, instead, is to vary such level within a same double-prism system, characterized by a fixed value of $n$. 

There are of course different ways to do so. A simple one is to consider that the two experimenters, who have to hit the prisms in the rolling experiments, may not just do it aimlessly, but trying to obtain the specific effect of having the two prisms rolling over a predetermined distance, thus producing a predetermined total angle of rotation. Of course, this will make a difference only when the two experimenters are executing the coincidence experiment $e_{ab}^{AB}$, as is clear from the fact that when they don't hit the two prisms together (in experiments $e_{ab'}^{AB}$ and $e_{a'b}^{AB}$), then because of the extreme low friction of the ``+'' faces (and the fact that, according to the protocol, they have to pull the knob firmly, i.e., hit the prism strongly) they will not succeed in altering the predetermined $(+,+)$ outcome.   

Now, let us assume that if the two players are successful in producing the chosen rotation in the coincidence experiment $e_{ab'}^{AB}$, then, considering the state in which the system was prepared, the outcome of the rolling experiment will be $(-1,+1)$ (or $(+1,-1)$). In general terms, we can characterize the ability of the two players in obtaining the desired effect by means of a continuous parameter $\epsilon\in [0,1]$, such that $\epsilon =1$ corresponds to the totally random situation, of maximum indeterminism, when the experimenters hit the two prisms without trying to obtain any specific result, and $\epsilon =0$ corresponds to the opposite situation where the two players are able to perfectly control their shot and therefore produce the $(-1,+1)$ (or $(+1,-1)$) outcome, without fail. In other terms, by varying $\epsilon$ from $1$ to $0$, we can decrease the degree of indeterminism in the measurement processes.

Following Ref.~\cite{AertsBroekaert}, we also want to introduce, in addition to the indeterminism parameter $\epsilon$, an additional continuous parameter $\rho\in[0,1]$, characterizing the connectedness of the two prisms (i.e., the strength of the correlations between the rectangular faces of the two prisms). A simple and natural way to do so is to consider the possibility that, sometimes, the rod can also detach and fall during the execution of the $e_{ab}^{AB}$ experiment. 

Let us simply assume that $\rho$ corresponds to the probability for the rod of remaining duly glued to the two prisms, when $e_{ab}^{AB}$ is executed. This means that $\rho=1$ corresponds to the case of maximum correlation, whereas $\rho=0$ to the case of absence of correlation. In other terms, by varying $\rho$ from $1$ to $0$, we can decrease the degree of connectedness (non-locality) of the two entangled prism-entities. Clearly, when the rod detaches, the two prism will necessarily end their respective roll by showing an upper face with the ``+'' symbol, and this means that the outcome $(+1,+1)$ is now also possible for the experiment $e_{ab}^{AB}$, with probability $1-\rho$. 

Assuming for simplicity a linear variation of the probabilities as a function of the ability parameter $\epsilon$, we can write for the four different outcomes of experiment $e_{ab}^{AB}$:
\begin{eqnarray}
\label{probabilities-epsilon-2}
{\cal P}_{ab}^{AB}(+1,+1)&=& 1-\rho, \nonumber\\
{\cal P}_{ab}^{AB}(-1,+1)&=& \rho +\rho \epsilon \left(\frac{2}{n} -1\right), \nonumber\\
{\cal P}_{ab}^{AB}(+1,-1)&=& \rho \epsilon\frac{2}{n},\nonumber\\
{\cal P}_{ab}^{AB}(-1,-1)&=& \rho \epsilon\frac{n-4}{n}.
\end{eqnarray}

Then, observing that the probabilities associated with the other three coincidence experiments are not directly affected by the values taken by $\epsilon$ and $\rho$ (this because, when the two prisms are rolled independently from one another, they will inevitably end their run by showing a ``+'' upper face), we obtain, after a short calculation: 
\begin{equation}
\label{I-epsilon-rho}
I=2+\rho\left[2(1-\epsilon) +\epsilon\frac{8}{n}\right].
\end{equation}

Visibly, (\ref{I-epsilon-rho}) generalizes (\ref{violation-n}), and in accordance with the analysis of the model in~\cite{AertsBroekaert}, we can observe the following. Only for the value $\rho=0$, corresponding to two totally disconnected prisms, BI is obeyed, and this means that it is the correlation between the two subsystems, i.e., the non-locality ingredient, which is really responsible for the violation. 

Also, we can see that when we increase $\epsilon$, i.e., the amount of uncertainty in the outcome of the experiments, we clearly also diminish the value of (\ref{I-epsilon-rho}), i.e., the degree of violation of BI, which means that, in accordance with~\cite{AertsBroekaert},  not only indeterminism (associated here to parameter $\epsilon$) doesn't contribute to the violation, but actually tends to reduce it. Clearly, the maximum violation, $I=4$, is obtained for $\rho =1$ and $\epsilon =0$, i.e., for the situation of perfect correlation and full predictability.

\section{Dependence on the initial state}
\label{Dependence on the initial state}

It is however important to observe that although inequality (\ref{violation-n}) is independent of the choice of the initial state of the double-prism system (which can either be $(-,+)$, $(+,-)$ or $(-,-)$), this is not anymore the case when the outcomes of the experiments are affected by the ability parameter $\epsilon$. Indeed, for a given rolling distance (i.e., for a given total angle of rotation) that the two experimenters will try to obtain, if we change the initial state of the system, then we will also change the expected outcome of the experiment, and therefore the value taken by BI. 

To see this, let us assume this time that the system is prepared in a state such that when the two experimenters can successfully impart to the double-prism the chosen total rotation angle, the final outcome is now $(-1,-1)$, instead of the previous $(-1,+1)$ (or $(+1,-1)$). Then, we obtain the following probabilities for the four outcomes of experiment $e_{ab}^{AB}$:
\begin{eqnarray}
\label{probabilities-epsilon-1-bis}
{\cal P}_{ab}^{AB}(+1,+1)&=&1-\rho, \nonumber\\
 {\cal P}_{ab}^{AB}(-1,+1)&=& \rho \epsilon \frac{2}{n},\nonumber\\
{\cal P}_{ab}^{AB}(+1,-1)&=& \rho \epsilon\frac{2}{n},\nonumber\\
{\cal P}_{ab}^{AB}(-1,-1)&=&\rho\left[1+ \epsilon\left(\frac{n-4}{n}-1\right)\right].
\end{eqnarray}

Observing once more that the probabilities associated to the other three coincidence experiments are not affected by the specific values taken by $\epsilon$ and $\rho$, we find after a short calculation: 
\begin{equation}
\label{I-epsilon-rho-bis}
I=2+\rho\epsilon\frac{8}{n}.
\end{equation}

As we can see, (\ref{I-epsilon-rho-bis}) differs sensibly from (\ref{I-epsilon-rho}), and the effect produced by a variation of parameter $\epsilon$ is now exactly opposite: an increase of the level of indeterminism (i.e., an increase of $\epsilon$) produces a strengthening of the violation of BI, and not a weakening of it. Also, the situation of full predictability ($\epsilon =0$) is not anymore associated to a maximal violation of the inequality, but to the non-violation of it!

\section{Discussion}
\label{Discussion}

We conclude with a few comments. First of all, it could be objected that, contrary to the model explored in~\cite{AertsBroekaert}, $\epsilon$ is not the only source of indeterminism in our model, considering that we have defined the parameter $\rho$ as a probability, and that the outcomes of the coincidence experiments $e_{ab}^{AB}$ clearly depend on whether the rod will detach or not during their execution. In other terms, $\rho$ also contributes to the degree of unpredictability of the outcomes of $e_{ab}^{AB}$. 

This is obviously true, but cannot alter our conclusion. Indeed, if we keep $\rho$ fixed, then an increase of $\epsilon$ does actually correspond to a global increase of the level of indeterminism in our model, which, according to (\ref{I-epsilon-rho}) and (\ref{I-epsilon-rho-bis}), can either reinforce or weaken the degree of the violation of BI, depending on the state in which the system was prepared.

In fact, our probabilistic description of non-locality in the model highlights an additional mechanism through which a variation of the amount of indeterminism can affect the value taken by BI, in a way that is not unique. Indeed, as regards the randomness incorporated in $\rho$, we can consider that the situation $\rho=1/2$ corresponds to the one of maximum uncertainty. Then, when from that value of maximum uncertainty we increase $\rho$, and therefore reduce the uncertainty, according to (\ref{I-epsilon-rho}) and (\ref{I-epsilon-rho-bis}) we increase the violation of BI, and this independently of the initial state of the system. But when starting from the same value $\rho=1/2$ we decrease $\rho$, also in this case we reduce the uncertainty, yet this time we decrease the violation of BI (which for the limit value $\rho=0$ is obeyed), independently of the initial state of the system.

Having said that, let us now explain why, when the level of indeterminism is increased in our model, by increasing $\epsilon$, we can either increase or decrease the violation of BI, depending on the initial state of the system. For simplicity, let us set $\rho=1$ (perfect correlation). Considering that in our model $E^{AB}_{ab'} = E^{AB}_{a'b'} =  E^{AB}_{a'b} =1$, independently of the value taken by $\epsilon$, we can write: $I= 2+|E^{AB}_{ab} - 1|$. Now, having assumed perfect correlation, it is clear that only three outcomes are possible for experiment $e_{ab}^{AB}$: $(-,-)$, $(+,-)$ and $(-,+)$. According to (\ref{expectation value}), outcome $(-,-)$ contributes positively to $ E^{AB}_{ab}$, whereas outcomes $(+,-)$ and $(-,+)$ contribute negatively to it. Therefore, if the outcome of experiment $e_{ab}^{AB}$ is predetermined, and corresponds to $(-,-)$, $E^{AB}_{ab}=1$ and BI is obeyed ($I=2$). On the other hand, if the predetermined outcome is  $(+,-)$, or $(-,+)$, $E^{AB}_{ab}=-1$ and BI is maximally violated ($I=4$).   

Assuming that we are in the situation where $E^{AB}_{ab}=1$, for $\epsilon=0$, then by increasing $\epsilon$ the system will start sometimes to also explore the outcomes $(+,-)$, or $(-,+)$, and since the latter contribute negatively to $E^{AB}_{ab}$, their possible selection will cause its value to diminish, thus producing an increase in the violation of BI. And of course, the situation is reversed when $E^{AB}_{ab}=-1$, for $\epsilon=0$.

The reason why Aerts et al. couldn't highlight in~\cite{AertsBroekaert} this double role played by indeterminism in coincidence experiments, is because the statistics of outcomes of their model does not depend on the specific state in which the system is prepared, but only on the relative orientation of the measuring apparatus (as is the case in spin measurements on singlet states, provided the direction of flight of the entangled pair is orthogonal to the directions of orientation of the Stern-Gerlach filters). Therefore, although they have studied a model which is more elaborated than ours (as meant to reproduce the same statistics as spin measurements on singlet states), it was actually too specific to fully elucidate the question of the role played by indeterminism in the violation of BI. 

To conclude, we briefly summarize our results. In agreement with~\cite{AertsBroekaert}, we have found that it is the aspect of non-locality, expressed by the connecting rod in our model, which produces the violation of BI, through the \emph{creation of correlations} between outcomes of experiments performed in coincidence.

In agreement with~\cite{AertsBroekaert}, we have also found that by increasing the indeterminism (increasing $\epsilon$), we can decrease the value taken by BI. However, we have also shown that this is not a general fact: depending on the state in which the system is prepared, an increase of the level of indeterminism is also capable of increasing the value taken by BI.

An additional source of indeterminism can also be envisioned, that was not considered in~\cite{AertsBroekaert}, associated to the possibility of actualizing different degrees of non-locality (connectedness) in a coincidence experiment. This possibility was described in our model by assuming that not only correlations between ``upper faces'' had to be considered as potential before a coincidence experiment, but also correlations between ``faces.'' 

When we do so, we find that if the amount of indeterminism associated to this additional level of potentiality is decreased, it can either increase or decrease the value of BI, depending on whether the decrease produces a strengthening or a weakening of the non-local aspect, respectively. As far as this author can judge, it is still an open experimental question to know if a quantum microscopic system in an entangled state, like a singlet state, has also a probability of ``breaking'' during a coincidence experiment (in the same way the rod in our macroscopic model has a probability of detaching), and produce in this way uncorrelated outcomes, instead of correlated ones.~\cite{Aerts5}

A last remark is in order. The description of the rolling/sliding behavior of the $n$-prisms must be understood, as we already said, in an idealized sense. We have never performed real experiments with systems of this sort, and therefore cannot guarantee that the way we have theoretically described their behavior, although reasonable, would be perfectly adequate from the perspective of a real experiment. But of course, this is not an essential point in the present analysis: what is important is that the idealized system we have considered behaves in a logical manner, according to coherent mechanisms. The question of how to exactly implement such behavior in real models, which can be subject to real experiments, is a technological issue which goes beyond the purely conceptual scope of the present note.

\end{document}